\documentclass[prd, onecolumn]{revtex4}

\usepackage{graphicx}
\usepackage{dcolumn}
\usepackage{bm}
\usepackage{wrapfig}
\usepackage{epsfig}
\usepackage{breqn}
\usepackage{amssymb}
\usepackage{subfigure}
 \usepackage{float}

\begin{document}


\title{Dynamics of a Bianchi Type I Model With a Concave Potential}

\author{Ikjyot Singh Kohli}
	\email{isk@mathstat.yorku.ca}
\affiliation{York University - Department of Mathematics and Statistics}
\author{Michael C. Haslam}
\email{mchaslam@mathstat.yorku.ca}
\affiliation{
York University - Department of Mathematics and Statistics
}

\date{March 18, 2017}

\begin{abstract}
In this paper, we study the dynamics of a Bianchi Type I potential in the presence of a concave potential of the form $V(\phi) = V_0 \left[1- \left( \frac{\phi}{\mu}\right)^n\right]$, where $V_0$ is a constant, and $\mu$ is a mass scale. We show that there are two classes of equilibrium points. The first class corresponds to $\phi \neq 0$, $\mu \neq 0$, $n = 0$, and $\dot{\phi} = 0$, which describe expanding and contracting de Sitter universes, for which the shear anisotropy is zero. We show that the expanding de Sitter universe is a local sink of the system, and therefore has associated to it a stable manifold. Thus, orbits will approach this point at late times. In other words, such a model is found to inflate and isotropize at late times as long as $n = 0$. The second class of equilibrium points corresponds to an expanding and contracting anisotropic universe. However, these points are found to emerge only when $n > 1$, $\phi = \dot{\phi} = 0$, which importantly implies that $V = V_0 < 0$ at this point in order to ensure that the square of the shear scalar, $\sigma^2$ is real. Therefore, such equilibrium points correspond to the ekpyrotic cosmological models. Further, we show that for $n = 2$, by Lyapunov's stability theorem, the expanding equilibrium point is asymptotically stable, while for $n > 2$, a two-dimensional stable manifold exists corresponding to the fact that for $n > 2$, such an equilibrium point represents a local sink of the system. Finally, we give a general condition for inflation to occur in this model in terms of the deceleration parameter, and show that the expanding ekpyrotic equilibrium point undergoes the phenomenon of anisotropic inflation if $-3\Lambda/5 < V_0 < 0$, where $\Lambda > 0$ is the cosmological constant.
\end{abstract}
\maketitle 

\section{Introduction}
In this paper, we are primarily concerned with analyzing the dynamics of a Bianchi Type I cosmological model containing a minimally coupled scalar field with a concave potential. Specifically, we consider a potential function of the form
\begin{equation}
V(\phi) = V_{0} \left[1 - \left(\frac{\phi}{\mu}\right)^n\right],
\end{equation}
where $n$ and $V_{0}$ are constants, and $\mu$ is a mass scale. Interestingly, such models include the so-called ``hilltop'' and new inflation models which are currently favoured by recent observations of the cosmic microwave background \cite{1674-1137-40-10-100001}.

The dynamics of Bianchi scalar field models have been considered in several contexts in the literature. Chimento and Labraga \cite{1998GReGr..30.1535C} examined the solutions to the Einstein-Klein-Gordon equations without a cosmological constant for an exponential potential in a Bianchi $VI_0$ metric. Billyard, Coley, and van den Hoogen \cite{1998PhRvD..58l3501B}  examined the dynamics of Bianchi models with a perfect fluid and a scalar field with an exponential potential. Billyard, Coley, van den Hoogen, Ib{\'a}{\~n}ez, and Olasagasti \cite{1999CQGra..16.4035B} studied the dynamics of Bianchi Type B cosmologies with matter with a barotropic fluid and a scalar field with exponential potential. In particular, they studied the asymptotic properties of the models both at early and late times, paying particular attention to whether the models isotropize (and inflate) in the future. Coley and Goliath \cite{2000PhRvD..62d3526C} analyzed the dynamics of various spatially homogeneous and closed cosmological models with a perfect fluid and a scalar field with an exponential potential. Aguirregabiria, Labraga, and Lazkoz \cite{2001gr.qc.....7009A} looked at the phenomenon of assisted inflation in Bianchi $VI_0$ cosmologies with multiple scalar fields with exponential potentials. Clarkson, Coley, and Quinlan \cite{2001PhRvD..64l2003C} analyzed the dynamics of Bianchi Type $VI_{0}$ cosmologies with a perfect fluid, a scalar field with an exponential potential, and a uniform cosmic magnetic field. Fay and Luminet \cite{2004CQGra..21.1849F} analyzed in detail the isotropization conditions for a Bianchi Type I model with scalar fields. Fay \cite{2005qcrt.conf..137F} studied the dynamics of different Bianchi models in the presence of a massive scalar field using the ADM Hamiltonian formalism. Lidsey \cite{2006CQGra..23.3517L} investigated the dynamics of a collapsing Bianchi Type IX model with a scalar field with a steep, negative exponential potential. Folomeev \cite{2007IJMPD..16.1845F} looked at the dynamics of a Bianchi Type I model with two interacting scalar fields. Aref'eva, Bulatov, and Vernov \cite{2010TMP...163..788A} performed a detailed stability analysis of isotropic solutions for two-field models in the Bianchi Type I metric. Rybakov, Shikin, Popov, and Saha \cite{2011IJTP...50.3421R} studied Bianchi Type I models and their ability to inflate and isotropize at late times in the presence of a nonlinear potential. Fadragas, Leon, and Saridakis \cite{2014CQGra..31g5018F} gave a very detailed analysis of the dynamics of Kantowski-Sachs, LRS Bianchi I and LRS Bianchi III models for a wide range of potentials. Shogin and Hervik \cite{2014CQGra..31m5006S} used dynamical systems methods to analyze Bianchi Type $VIII$ cosmologies with a tilted fluid undergoing diffusion on a scalar field. Ribas, Samojeden, Devecchi, and Kremer \cite{2015PhyS...90j5001R} studied the dynamics of a Bianchi Type I model in the presence of a Yukawa potential. 

In this work, we look at the dynamics of a Bianchi Type I model with a concave potential that has not been to the best of the authors' knowledge, previously studied in the literature. Such an analysis can be useful, since the early universe may have had strong anisotropies \cite{elliscosmo}, combined with the fact that recent data favours such concave potentials in inflationary scenarios \cite{1674-1137-40-10-100001}. Throughout, we use units where $8\pi G = c = 1$.

\section{The Dynamical Equations}
We will employ the orthonormal frame formalism as in \cite{ellismac} to obtain the Einstein field equations as a coupled system of first-order, ordinary differential equations as follows. Since the shear tensor is trace-free, and can be written in the form $\sigma_{ab} = \text{diag}(\sigma_{11}, \sigma_{22}, \sigma_{33})$, we will define the shear components of the Bianchi Type I model as:
\begin{equation}
\label{eq:sheardef}
\sigma_{+} = \frac{1}{2}\left(\sigma_{22} + \sigma_{33}\right), \quad \sigma_{-} = \frac{1}{2\sqrt{3}} \left(\sigma_{22} - \sigma_{33}\right).
\end{equation}
From this, we then have that
\begin{equation}
\label{eq:shearscalar}
\sigma^2 \equiv \frac{1}{2}\sigma_{ab}\sigma^{ab} = 3\left(\sigma_{+}^2 + \sigma_{-}^2\right).
\end{equation}

Since the Bianchi type I model is also spatially flat, the shear propagation equations take the simple form:
\begin{equation}
\label{eq:shearprop1}
\dot{\sigma}_{\pm} = -\theta \sigma_{\pm},
\end{equation}
where $\theta$ is the expansion scalar.

The other dynamical equations are the Raychaudhuri equation:
\begin{equation}
\label{eq:raych1}
\dot{\theta} = -\frac{1}{3}\theta^2 - \sigma_{ab}\sigma^{ab} -\frac{1}{2}\left(\rho + 3p\right) + \Lambda,
\end{equation}
and the Friedmann equation:
\begin{equation}
\label{eq:fried1}
\frac{1}{3}\theta^2 = \frac{1}{2}\sigma_{ab} \sigma^{ab} + \rho + \Lambda.
\end{equation}

In Eqs. Eq. (\ref{eq:raych1}) and Eq.(\ref{eq:fried1}), $\rho$ and $p$ are the matter energy density and pressure respectively, while $\Lambda$ is the cosmological constant. As for the matter content in our model, we are assuming a scalar field $\phi$ that is minimally coupled to gravity, such that the energy-momentum tensor takes the form \cite{elliscosmo}
\begin{equation}
T^{ab} = \nabla^{a} \phi \nabla^{b} \phi - \left(\frac{1}{2} \nabla_{c} \phi \nabla^{c} \phi + V(\phi)\right)g^{ab}.
\end{equation}
In an orthonormal frame formalism, we will assume that the four-velocity of the $\phi$-field is given as
\begin{equation}
u_{a} = -\frac{1}{\dot{\phi}} \nabla_{a} \phi,
\end{equation}
so that $T_{ab}$ is in the form of a perfect fluid, and we have that
\begin{equation}
\rho = \frac{1}{2} \dot{\phi}^2 + V(\phi), \quad p = \frac{1}{2}\dot{\phi}^2 - V(\phi).
\end{equation}
Further, the conservation condition on $T^{ab}$, namely that $\nabla_{b} T^{ab} = 0$ gives the Klein-Gordon equation which describes the dynamics of the $\phi$-field:
\begin{equation}
\label{eq:kg1}
\ddot{\phi} + \theta \dot{\phi} + V'(\phi) = 0.
\end{equation}

As mentioned above, for this model, we would like to consider a potential in the form of
\begin{equation}
\label{eq:potential}
V(\phi) = V_{0} \left[1 - \left(\frac{\phi}{\mu}\right)^n\right],
\end{equation}
where $n$ and $V_{0}$ are constants, and $\mu$ is a mass scale. 

Together, Eqs. (\ref{eq:shearprop1}), (\ref{eq:raych1}), (\ref{eq:fried1}), and (\ref{eq:kg1}) comprise the dynamics of our cosmological model. To employ a dynamical systems approach, we further let $f \equiv \dot{\phi}$, and upon using Eqs. (\ref{eq:shearscalar}) and (\ref{eq:fried1}) to decouple the shear variables, we obtain a dynamical system as follows:
\begin{eqnarray}
\label{eq:dyn1}
\dot{\theta} &=& -2f^2 - \theta^2 + 3\Lambda + V_{0} \left[1 - \left(\frac{\phi}{\mu}\right)^n\right], \\
\label{eq:dyn2}
\dot{\phi} &=& f, \\
\label{eq:dyn3}
\dot{f} &=& -f \theta + \frac{n V_{0} \left( \frac{\phi}{\mu}\right)^n}{\phi}.
\end{eqnarray}

\section{A Stability Analysis}
With the dynamical equations (\ref{eq:dyn1}-\ref{eq:dyn3}) in hand, we can now perform a stability analysis of the equilibrium points of the system. There are two classes of equilibrium points which we now describe.

\subsection{Equilibrium Points 1 and 2}
The first class of equilibrium points occurs for $\phi \neq 0$, $\mu \neq 0$, $n = 0$, and $f=0$. At this point, we have that $\theta = \pm \sqrt{3} \sqrt{\Lambda}$, which implies from Eq. (\ref{eq:fried1}) that $\sigma^2 = 0$. These equilibrium points represent expanding and contracting de Sitter universes respectively, which, in the expansion case (the positive root of $\theta$) represents an inflationary epoch of this model. We will denote these points by $P_{1}$ and $P_{2}$ respectively. The eigenvalues of the Jacobian matrix associated to Eqs. (\ref{eq:dyn1}-\ref{eq:dyn3}) at $\theta = \sqrt{3} \sqrt{\Lambda}$ are given by
\begin{equation}
\label{eq:eigs1}
\lambda_{1} = 0, \quad \lambda_{2} = -2\sqrt{3} \sqrt{\Lambda}, \quad \lambda_{3} = -\sqrt{3} \sqrt{ \Lambda}.
\end{equation}
In this case, there is a two-dimensional stable manifold (since $\Lambda > 0$), and a one-dimensional center manifold because of the zero eigenvalue. Even though there is a zero eigenvalue, by the invariant manifold theorem, this equilibrium point is normally hyperbolic \cite{ellis}. Hence, the local stability is determined by the signs of the real part of the non-zero eigenvalues, $\lambda_{2}$ and $\lambda_{3}$. Therefore, this equilibrium point is a local sink of the dynamical system.

The eigenvalues of the Jacobian matrix associated to Eqs. (\ref{eq:dyn1}-\ref{eq:dyn3}) at $\theta = -\sqrt{3} \sqrt{\Lambda}$ are given by
\begin{equation}
\label{eq:eigs2}
\lambda_{1} = 0, \quad \lambda_{2} = 2\sqrt{3} \sqrt{\Lambda}, \quad \lambda_{3} = \sqrt{3} \sqrt{ \Lambda}.
\end{equation}
In this case, there is a two-dimensional unstable manifold (since $\Lambda > 0$), and a one-dimensional center manifold because of the zero eigenvalue. Even though there is a zero eigenvalue, by the invariant manifold theorem, this equilibrium point is normally hyperbolic \cite{ellis}. Hence, the local stability is determined by the signs of the real part of the non-zero eigenvalues, $\lambda_{2}$ and $\lambda_{3}$. Therefore, this equilibrium point is a source of the dynamical system.

\subsection{Equilibrium Points 3 and 4}
There are two interesting equilibrium points that emerge from this dynamical system that occur for when $n$, the exponent in Eq. (\ref{eq:potential}) is such that $n > 1$. These equilibrium points are given as:
\begin{equation}
\label{eq:eqpoint34}
\theta = \pm \sqrt{3\Lambda - V_{0}}, \quad \phi = 0, \quad f = 0.
\end{equation}
We will denote these points by $P_{3}$ and $P_{4}$ respectively.
From Eq. (\ref{eq:fried1}), at these points, the square of the shear scalar, $\sigma^2$ takes the value
\begin{equation}
\label{eq:shear1}
\sigma^2 = -\frac{4}{3}V_{0},
\end{equation}
which is only valid if $V_{0} < 0$. Looking at the potential function described by Eq. (\ref{eq:potential}), one sees that $V_{0} < 0$ implies in fact that $V(\phi) < 0$ at this equilibrium point. Interestingly, this cosmological model has associated to it an equilibrium point that describes the so-called ekpyrotic model \cite{elliscosmo, 2001PhRvD..64l3522K}, that is, allows for negative potential energy.

The eigenvalues associated with the positive root of $\theta$ are found to be:
\begin{eqnarray}
\label{eq:eigs3}
\lambda_{1} = -2 \sqrt{3 \Lambda -V_0}, \quad \lambda_{2,3} = \frac{1}{2} \left(-\sqrt{3 \Lambda +\left(\frac{8}{\mu^2}-1\right) V_0} \pm \sqrt{3 \Lambda -V_0}\right), &\quad& n = 2, \nonumber \\
\label{eq:eigs4}
\lambda_{1} = -2 \sqrt{3 \Lambda -V_0}, \quad \lambda_{2} = -\sqrt{3 \Lambda -V_0}, \quad \lambda_{3} = 0, &\quad& n > 2.
\end{eqnarray}
It can be easily shown that this point is never a saddle or source. Further, since all of the eigenvalues are negative at this point for $n=2$, by Lyapunov's stability theorem \cite{arnolddyn}, this point is asymptotically stable. That is, it is Lyapunov stable, and, in addition, all solutions with initial conditions sufficiently close to this point tend to this point as $t \to \infty$.

%

For the case when $n > 2$, we clearly have a one-dimensional center manifold from Eq. (\ref{eq:eigs4}). However, the stability can be analyzed from the signs of the real parts of the other eigenvalues. Indeed, in this case, the other two eigenvalues are always negative. Hence, this point is also a local sink for the case where $n > 2$.

The remaining equilibrium point corresponds to the negative root of $\theta$ in Eq. (\ref{eq:eqpoint34}). The eigenvalues associated with this equilibrium point are found to be
\begin{eqnarray}
\label{eq:eigs5}
\lambda_{1} = 2 \sqrt{3 \Lambda -V_0}, \quad \lambda_{2,3} = \frac{1}{2} \left(\sqrt{3 \Lambda -V_0} \pm \sqrt{3 \Lambda +\left(\frac{8}{\mu^2}-1\right) V_0}\right), &\quad& n =2, \\
\label{eq:eigs6}
\lambda_{1} = 2 \sqrt{3 \Lambda - V_0}, \quad \lambda_{2} = 0, \quad \lambda_{3} = \sqrt{3 \Lambda -V_0}, &\quad& n > 2.
\end{eqnarray}

One can see that for the case when $n=2$, this equilibrium point is always a source of the dynamical system.
As can be confirmed by looking at the signs of the real parts of the eigenvalues, this point is never a local sink or saddle of the system.


For the case when $n > 2$, we clearly have a one-dimensional center manifold from Eq. (\ref{eq:eigs6}). However, the stability can be analyzed from the signs of the real parts of the other eigenvalues. Hence, this point is always a source of the system. 

In summary, we have expanding and contracting de Sitter universe equilibrium points when $n = f = 0$, where $\sigma^2 = 0$. We also have for when $n > 1$, expanding and contracting anisotropic universes that only emerge from the dynamics when $V_0 < 0$. Further, the expanding solutions in this case at least for $n=2$ are asymptotically stable. For $n > 2$, the expanding anisotropic solutions are local sinks of the system, while the contracting solutions are sources. In this way, it is interesting to note from a dynamical systems perspective, that there is a heteroclinic orbit joining the expanding and contracting de Sitter universe equilibrium points. There is also a heteroclinic orbit that emerges when $V_0 < 0$ that joins the expanding and contracting anisotropic solutions. 

\section{Conditions for Inflation}
In this section, we describe the conditions for inflation with respect to the equilibrium points found above. Namely, it is convenient to describe the condition for inflation in terms of the following parameter, $\epsilon$:
\begin{equation}
\label{eq:infcondition1}
\epsilon \equiv -\frac{3 \dot{\theta}}{\theta^2} \ll 1.
\end{equation}
Upon using Eq. (\ref{eq:dyn1}) and the general definition for the evolution of the expansion scalar,
\begin{equation}
\label{eq:thetadef}
\dot{\theta} = -\frac{1}{3}\left(1+q\right)\theta^2,
\end{equation}
where $q$ is the deceleration parameter, together, these equations imply that inflation will occur whenever $q \ll 0$. In fact, upon using Eqs. (\ref{eq:infcondition1}) and (\ref{eq:thetadef}), we obtain the following condition for inflation to occur:
\begin{equation}
\label{eq:infcond2}
q = \frac{6 f^2+2 \theta ^2-9 \Lambda +3 V_0 \left(\left(\frac{\phi }{\mu }\right)^n-1\right)}{\theta ^2} \ll 0.
\end{equation}

For the expanding de Sitter universe equilibrium point, $P_1$ one gets from Eq. (\ref{eq:infcond2}) that $q = -1 < 0$. Further, since we showed above that $P_1$ is a local sink of the system, this model undergoes inflation and isotropizes at late times for $n=0$ only. In other words, because $q = -1$ at this point, this equilibrium point represents an inflationary epoch. This seems to be similar to results obtained in \cite{1993PhRvD..48.4662A} where the dynamics of a Bianchi Type I model were considered in the presence of a scalar field potential proportional to $\exp(k \phi)$. They found that such a model isotropized for $k < \sqrt{2}$, but not for $k > \sqrt{2}$.

For the expanding anisotropic universe equilibrium point, $P_3$ one gets from Eq. (\ref{eq:infcond2}) that
\begin{eqnarray}
q &=& \frac{3 \Lambda +5 V_0}{V_0-3 \Lambda}, \quad \forall \ n > 1.
\end{eqnarray}
Therefore, inflation will occur when
\begin{equation}
\label{eq:ineq1}
-\frac{3 \Lambda}{5} < V_0 < 0,
\end{equation}
where we have enforced the condition that $V < 0$ at this equilibrium point to ensure that $\sigma^2 > 0$ per our analysis of the equilibrium points above. We showed above that $P_3$ is asymptotically stable for $n=2$, and is a local sink of the system for $n > 2$. Further, strictly under the condition (\ref{eq:ineq1}), this equilibrium point represents an inflationary epoch of the universe model under consideration. Interestingly though, it is an example of where the universe undergoes anisotropic inflation, that is, despite inflation occurring, the anisotropy does not vanish at late times. This is because, at this equilibrium point, $\sigma^2$ is non-zero, and it is a local sink of the system for $n > 2$, and asymptotically stable for $n=2$. This is a further example of so-called anisotropic inflation where it has been discovered previously that certain Bianchi Type I models can undergo inflation, but this inflation would be anisotropic \cite{2010PhRvD..81f3528G}. 

\section{Conclusions}
In this paper, we studied the dynamics of a Bianchi Type I potential in the presence of a concave potential. We showed that there are two classes of equilibrium points. The first class corresponded to $\phi \neq 0$, $\mu \neq 0$, $n = 0$, and $f = \dot{\phi} = 0$, which described expanding and contracting de Sitter universes, for which the shear anisotropy was zero. We showed that the expanding de Sitter universe was a local sink of the system, and therefore has associated to it a stable manifold. Thus, orbits will approach this point at late times. In other words, such a model was found to inflate and isotropize at late time as long as $n = 0$. The second class of equilibrium points corresponded to an expanding and contracting anisotropic universe. However, these points were found to emerge only when $n > 1$, $\phi = f = \dot{\phi} = 0$, which importantly implied that $V = V_0 < 0$ at this point in order to ensure that the square of the shear scalar was real. Therefore, such equilibrium points correspond to the ekpyrotic cosmological models. Further, we showed that for $n = 2$, by Lyapunov's stability theorem, the expanding anisotropic equilibrium point was asymptotically stable, while for $n > 2$, a two-dimensional stable manifold exists corresponding to the fact that for $n > 2$, this equilibrium point represented a local sink of the system. Finally, we gave a general condition for inflation to occur in this model in terms of the deceleration parameter, and showed that the ekpyrotic equilibrium points undergo the phenomenon of anisotropic inflation if $-3\Lambda/5 < V_0 < 0$, where $\Lambda > 0$ is the cosmological constant.

\newpage 
\bibliography{sources}

\end{document}